\documentclass[aps,prl,twocolumn,preprintnumbers,superscriptaddress,nofootinbib]{revtex4-1}
\usepackage[utf8]{inputenc}
\usepackage{psfrag}
\usepackage{amssymb}
\usepackage{amsmath}
\usepackage{epstopdf}
\usepackage{color}
\usepackage{graphicx}
\usepackage{subcaption}
\usepackage{booktabs}

\usepackage{lmodern}
\usepackage{hyperref}
\hypersetup{colorlinks=true,linkcolor=red,anchorcolor=blue,citecolor=blue, filecolor=blue,urlcolor=red,bookmarksnumbered=true,
pdfview=FitB
}

\begin{document}
\title{Structure of lightest nuclei in the visible Universe}
\author{Satvir~Kaur}
\email{satvir@impcas.ac.cn}
\affiliation{Institute of Modern Physics, Chinese Academy of Sciences, Lanzhou 730000, China}
\affiliation{University of Chinese Academy of Sciences, Beijing 100049, China}
\author{Chandan~Mondal}
\email{mondal@impcas.ac.cn} 
\affiliation{Institute of Modern Physics, Chinese Academy of Sciences, Lanzhou 730000, China}
\affiliation{University of Chinese Academy of Sciences, Beijing 100049, China}
\author{Xingbo~Zhao}
\email{xbzhao@impcas.ac.cn} 
\affiliation{Institute of Modern Physics, Chinese Academy of Sciences, Lanzhou 730000, China}
\affiliation{University of Chinese Academy of Sciences, Beijing 100049, China}
\affiliation{Advanced Energy Science and Technology Guangdong Laboratory, Huizhou, Guangdong 516000, China}
\author{Chueng-Ryong Ji}
\email{crji@ncsu.edu} 
\affiliation{Department of Physics, North Carolina State University, Raleigh, NC 27695-8202, USA}

\begin{abstract}
The simplest atomic nucleus, deuteron, provides key insights into the strong nuclear interactions among quarks and gluons that shape the visible universe. We present the first attempt to calculate the internal structure of the deuteron by incorporating hidden-color degrees of freedom, modeling it as an effective mixture of singlet-singlet and octet-octet color clusters beyond the traditional proton-neutron picture.
By employing the separation of variables for the light-front two-cluster bound-state equation, we explore how these hidden color correlations shape both its spin and electromagnetic structure. We incorporate the transverse and longitudinal dynamics by two Schr\"odinger-like equations, namely the light-front holography and the 't Hooft equation, respectively.
Our predictions of the electromagnetic form factors and structure functions, including tensor-polarized function, align well with experimental data, offering insights into the partonic structure of the deuteron. Its tensor property could pave the way for a new era in spin physics, guiding future experimental investigations.
\end{abstract}

\maketitle

{\it Introduction}.---The universe is composed predominantly of dark matter and dark energy, with ordinary visible matter, such as stars, planets, and galaxies, accounting for less than 5\% of its total content~\cite{Anderle:2021wcy}. Despite its scarcity, understanding the visible matter remains incomplete. While atoms, the fundamental building blocks of matter, consist of electrons orbiting a dense nucleus, these nuclei, in turn, are made up of nucleons which themselves are composite particles formed by quarks and gluons bound together by strong interaction, as described by quantum chromodynamics (QCD)~\cite{Callan:1977gz}. However, 
the QCD description of nuclei as multi-quark systems is not yet well understood due to the complicated dynamics among color degrees of freedom. In particular, we notice more than an order of magnitude increase of ``hidden-color" states~\cite{Ji:2014cda,Bakker:2014cua} in the nine-quark system compared to the six-quark system, i.e. $\frac{42-1}{5-1}=\frac{41}{4} >  10$.
Thus, we pay attention to a remarkable proliferation of the hidden color degrees of freedom in multi-quark systems.

Our study in the present work focuses on the deuteron, the simplest multi-quark system or atomic nucleus, as it provides a role model for studying the transition between the nucleon-meson description of nuclear physics and the fundamental quark-gluon picture of QCD. Traditionally, the deuteron is described as a loosely bound state of a proton and a neutron, where meson exchange dominates at large distances (low energies). However, 
at short distances from the more fundamental QCD perspectives, the deuteron is a six-quark system, where non-trivial color dynamics play a crucial role in its internal structure~\cite{Bakker:2014cua}.

In QCD, quarks must combine to form an overall color-neutral state. Within deuteron, the two nucleons can mix their internal color configurations, allowing quarks to rearrange into color-octet clusters. These octet-octet combinations, known as ``hidden-color'' states, preserve overall color neutrality despite each cluster carrying color. Such configurations, predicted by SU(3) color symmetry, are not captured by traditional nucleon-based description. Hidden colors are expected to play a key role in determining the deuteron’s short-distance structure, where a full quark-gluon treatment becomes essential~\cite{Matveev:1977xt, Matveev:1977ha, Hogaasen:1979qa, Brodsky:1983vf, Farrar:1991qi, Bashkanov:2013cla, Miller:2013hla, Bakker:2014cua}.

The deuteron, a spin-1 system, exhibits unique features such as tensor forces, a magnetic dipole moment, and an electric quadrupole moment. These properties make it a valuable system for exploring spin-dependent aspects of QCD in nuclei. Its electromagnetic properties have been studied extensively using various theoretical models~\cite{Gutsche:2016lrz, Gutsche:2015xva, Sun:2016ncc, Shi:2022blm, Huseynova:2022tok, Marcucci:2015rca, Khachi:2023iqp, Chemtob:1974nf, Blankenbecler:1971xa, Piller:1995mf, Karmanov:1994ck, Karmanov:1991fv} and through global analyses of elastic electron-deuteron scattering data~\cite{JLABt20:2000qyq, Sick:1998cvq, Zhang:2011zu, Nikolenko:2003zq}; see Refs.~\cite{Marcucci:2015rca, Garcon:2001sz}. 

\begin{figure}[h!]
	\centering
	\includegraphics[scale=0.19]{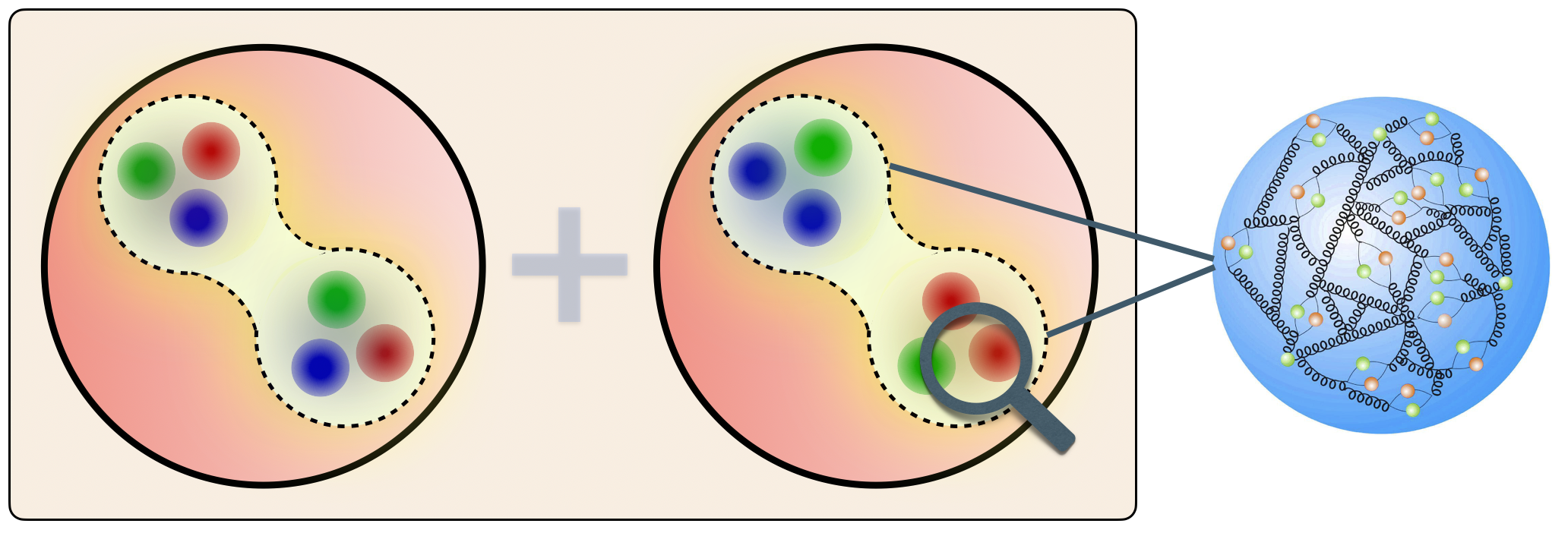}
	\caption{Visual representation of deuteron in our approach as an effective mixture of singlet-singlet and octet-octet states.
    }
	\label{fig:deuteron}
\end{figure}

However, the internal spin structure of the deuteron at the quark-gluon level is not fully understood. While its total spin is known, the roles of quark orbital motion, gluons, and sea quarks are still unclear. A key observable in this context is the tensor-polarized structure function $b_1$, specific to spin-1 systems and expected to vanish if the deuteron were merely a proton-neutron bound state~\cite{Kumano:2024fpr}. The HERMES experiment measured a nonzero $b_1$~\cite{HERMES:2005pon}, indicating a more complex internal structure.

Upcoming experiments at Jefferson Lab~\cite{ProposalJLab2023,Poudel:2025tac} and FermiLab~\cite{AbilityFermiLabE1039, Keller:2022abm} aim to explore this further. Discrepancies between theoretical predictions~\cite{Hoodbhoy:1988am, Khan:1991qk, Edelmann:1997ik, Cosyn:2017fbo, Kumano:2024fpr} and existing data point to gaps in our understanding of nuclear spin structure. Future experiments at the Electron-Ion Colliders (EICs) in the USA and China are expected to shed light on deuteron quark-gluon dynamics~\cite{Accardi:2012qut, Anderle:2021wcy}.

In this Letter, we investigate the deuteron as two interacting three-quark clusters (see Fig.~\ref{fig:deuteron})
which may be viewed not as two colorless clusters in low energy but as a mixture with two colorful clusters in
high energy. 
While the singlet-singlet state dominates at very low energy scales, perturbative QCD predicts that at extremely high energy scales, the octet-octet configuration contributes up to 80\%, with only 20\% from the singlet-singlet state~\cite{Ji:1985ky}. Thus, we consider an effective mixture of singlet-singlet and octet-octet color structures as our two-cluster degrees of freedom. This is far different from the traditional nucleon-nucleon picture which 
assumes a purely singlet-singlet color configuration.

As its first trial, we intermediate the two extreme color configurations of singlet-singlet vs.~octet-octet although their precise proportionality is apriori unknown. From our model computation, however, relatively large binding energy (B.E.) at the QCD scale of $\Lambda_{\text{QCD}}\sim \mathcal{O}(100~\text{MeV})$, which significantly exceeds the typical deuteron B.E.~$\sim$~2.2~MeV, would suggest that the shorter-distance dynamics is probed beyond the conventional neutron-proton picture. Assuming that the hidden-color degrees of freedom obey our proposed model equations of motion, we construct light-front wavefunctions (LFWFs) that describe the internal structure of the deuteron and compute the corresponding B.E. to gauge the involvement of hidden color degrees of freedom. For simplicity, only the $L = 0$ ($S$-wave) spin structure is included to isolate hidden-color effects. This approach provides a tractable framework to study the significance of QCD dynamics manifest in nuclear binding.

{\it Formalism}.---We investigate the motion of relativistic constituents inside the deuteron in both transverse and longitudinal momentum planes. The transverse dynamics are obtained by solving the holographic Schr\"odinger-like equation~\cite{Brodsky:2014yha}, while the longitudinal dynamics and non-zero constituent masses are incorporated through the 't Hooft equation~\cite{tHooft:1974pnl}, formulated in \((1+1)\)-dim QCD in the \(N_c \gg 1\) limit. Together, these equations provide a comprehensive description of QCD bound states~\cite{Ahmady:2021lsh, Chabysheva:2012fe, Sharma:2023njj, Ahmady:2021yzh, Ahmady:2022dfv, Gurjar:2024wpq}. 

Notably, our approach draws a natural analogy to the well-known Schr\"odinger equation for the hydrogen atom, where the wavefunction is separated into radial and orbital components. 
This separation elegantly distinguishes between energy contributions from radial motion and orbital angular momentum.

For the deuteron, we analogously split its energy structure and also the wavefunction into longitudinal and transverse components.
This connection not only provides a valuable perspective for analyzing the internal dynamics of the deuteron, but also offers a bridge between traditional non-relativistic quantum mechanics and the relativistic framework of light-front (LF) dynamics.

As such, we assume the dynamics of the clusters inside the deuteron are governed by the transverse and longitudinal LF Schr\"odinger equations:
\begin{equation}
 \left(-\frac{{\rm d}^2}{{\rm d} \zeta^2}+\frac{4L^2-1}{4 \zeta^2}+U_\perp(\zeta)\right)\phi(\zeta)= M_\perp^2 \phi(\zeta),
 \label{Eq:transverse}
\end{equation}
and
\begin{equation}
\left(\frac{m_\mathcal{C}^2}{z(1-z)}+U_\parallel(z)\right)\chi(z) = M^2_\parallel \chi(z),
\label{Eq:longitudinal}
\end{equation}
respectively. Here, $\phi(\zeta)$ represents the transverse wavefunction with $\boldsymbol{\zeta}=\sqrt{z(1-z)}{\bf b}_\perp$ being related to the separation between the clusters (${\bf b}_\perp$), and $X(z)=\sqrt{z(1-z)}\chi(z)$ defines the longitudinal wavefunction. The longitudinal momentum fraction carried by the active cluster is represented with $z$. The mass squared eigenvalue is given by $M^2 =M_\perp^2+M_\parallel^2$~\cite{Ahmady:2021lsh}.

The transverse potential in Eq.~\eqref{Eq:transverse}, derived from the holographic QCD approach~\cite{Brodsky:2014yha}, is given by
\begin{equation}
	U_\perp(\zeta)=\kappa^4 \zeta^2 + 2\kappa^2(J-1),
\label{Eq:U-LFH}
\end{equation}
with $J=L+S$ being the total angular momentum of the system, and
the longitudinal potential in Eq.~\eqref{Eq:longitudinal} is defined by 't Hooft as~\cite{tHooft:1974pnl},
\begin{equation}
U_\parallel(z)\chi(z)=\frac{g^2}{\pi} \mathcal{P}\int {\rm d}Z \frac{\chi(z)-\chi(Z)}{(z-Z)^2},
\label{eq:tHooft}
\end{equation}
where $\kappa$ and $g$ describe the strength of confinement in transverse and longitudinal directions, respectively. The symbol $\mathcal{P}$ in Eq.~\eqref{eq:tHooft} represents the principal value.

\begin{figure}[h!]
	\centering
	\includegraphics[scale=0.3]{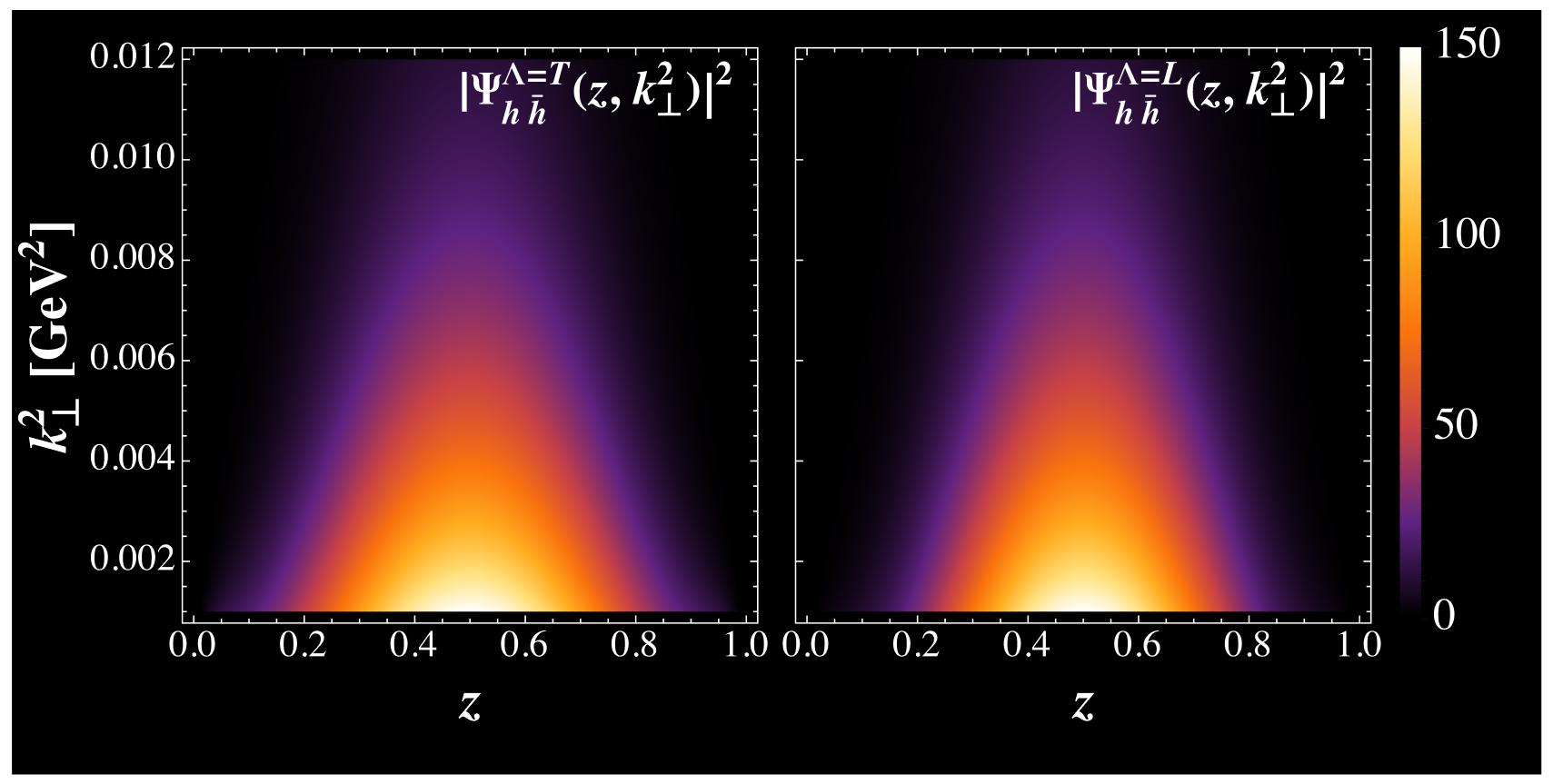}
	\caption{Probability distributions of transverse ($\Lambda=T$) and longitudinal ($\Lambda=L$) components of LFWFs.}
	\label{fig:LFWFs}
\end{figure}

The complete wavefunction is defined as 
\begin{equation}
	\Psi (z, \zeta, \theta)= \frac{\phi (\zeta)}{\sqrt{2\pi \zeta}}	 e^{\iota L \theta}
 X(z).
	\label{Eq:full-mesonwf}
\end{equation}
We construct the two-cluster spin wavefunction using the Melosh transformation~\cite{Melosh:1974cu}, which relates the instant-form spinors to their LF counterparts~\cite{Ji:1992yf}. The full LFWFs in momentum space are then given by~\cite{Forshaw:2012im}
\begin{align}
\Psi^\Lambda_{h,\bar{h}}(z,{\bf k}_\perp)&=\frac{\bar{v}_{\bar{h}}\left(\bar{z},-{\bf k}_\perp\right)}{\sqrt{\bar{z}}}\gamma^\mu \epsilon_\mu^\Lambda \frac{u_h(z,{\bf k}_\perp)}{\sqrt{z}} \Psi(z, {\bf k}_\perp),
\label{eq:spin-LFWFs}
\end{align}
where $\bar{z}=(1-z)$; $\epsilon^\Lambda_\mu$ is the polarization vector of the spin-1 system;  $u_h$ and $v_{\bar{h}}$ are the LF spinors of the clusters; and $\Psi(z, {\bf k}_\perp)$ is the Fourier transform of the spatial wavefunction $\Psi(z,{\bf b}_\perp)$ given in Eq.~\eqref{Eq:full-mesonwf}.

{\it Observables}.---The LF holographic equation and the 't~Hooft equation together capture 3D QCD confinement, providing LFWFs 
that successfully describe deuteron observables, such as electromagnetic form factors (EMFFs) and partonic distributions.

These predictions involve only three parameters: cluster mass $m_{\mathcal{C}}$, transverse scale $\kappa$, and longitudinal scale $g$. They are fixed by fitting the deuteron mass $M_{D} = 1.875 \pm 0.185$ GeV and the low-$Q^2$ behavior ($Q^2 \leq 0.5$ GeV$^2$) of $G_C$ and $G_M$, achieving $\chi^2/\text{d.o.f.} = 0.98$. The best-fit values are $\lbrace{m_{\mathcal{C}}, \kappa, g}\rbrace = \lbrace{0.838 \pm 0.083, 0.13 \pm 0.013, 0.50 \pm 0.05}\rbrace$ GeV.

Notably, the B.E. of the deuteron in our framework is attained as $\sim$$\,200$~MeV, calculated using $\text{B.E.} \approx  M_D-2m_{\mathcal{C}}$. This relatively large value suggests a significant contribution from hidden-color configurations, particularly octet-octet states. While the precise decomposition into singlet-singlet and octet-octet components cannot be identified in the present stage of our analysis, the result clearly points to a non-negligible hidden-color component beyond the conventional nucleon-nucleon picture.

Figure~\ref{fig:LFWFs} shows that, at zero relative longitudinal momentum fraction, the probability of finding a cluster is higher in a longitudinally polarized deuteron than when it is transversely polarized. However, the distribution is broader for transverse polarization, indicating a wider spread in cluster probability in momentum space.

	\begin{figure*}[hbt!]
	\centering
\includegraphics[width=0.325\textwidth]{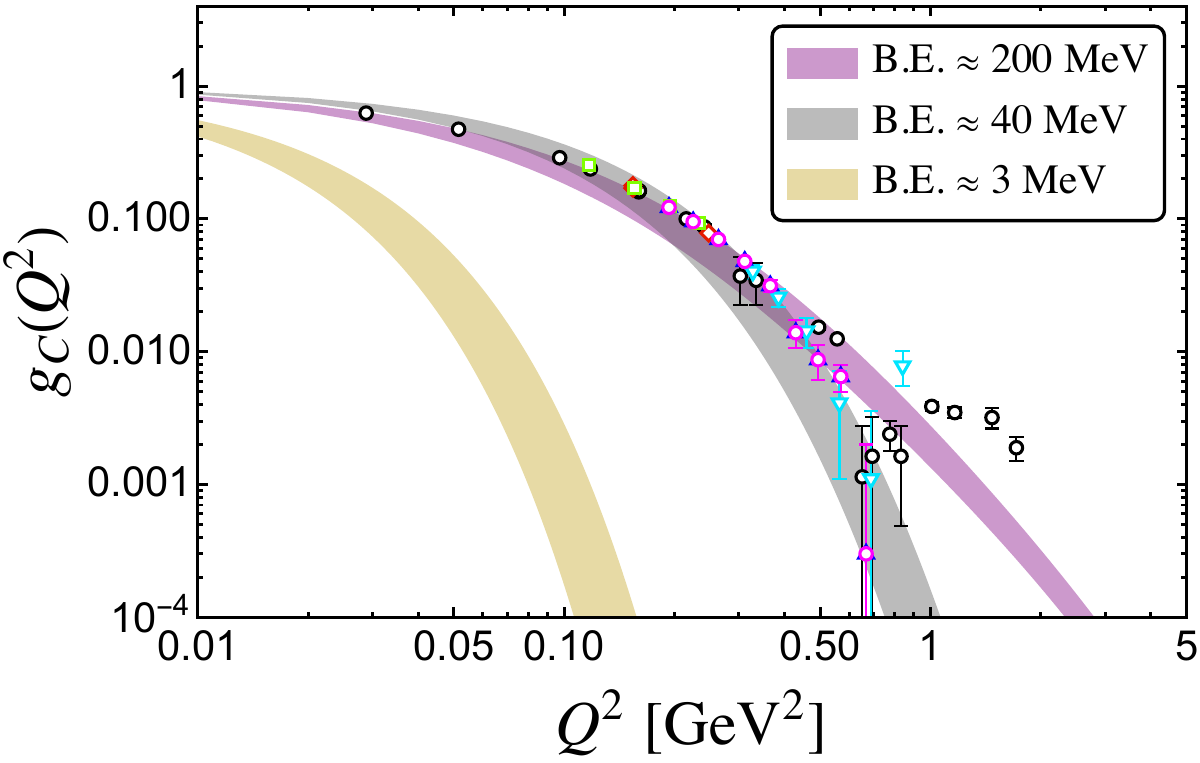}
    \includegraphics[width=0.325\textwidth]{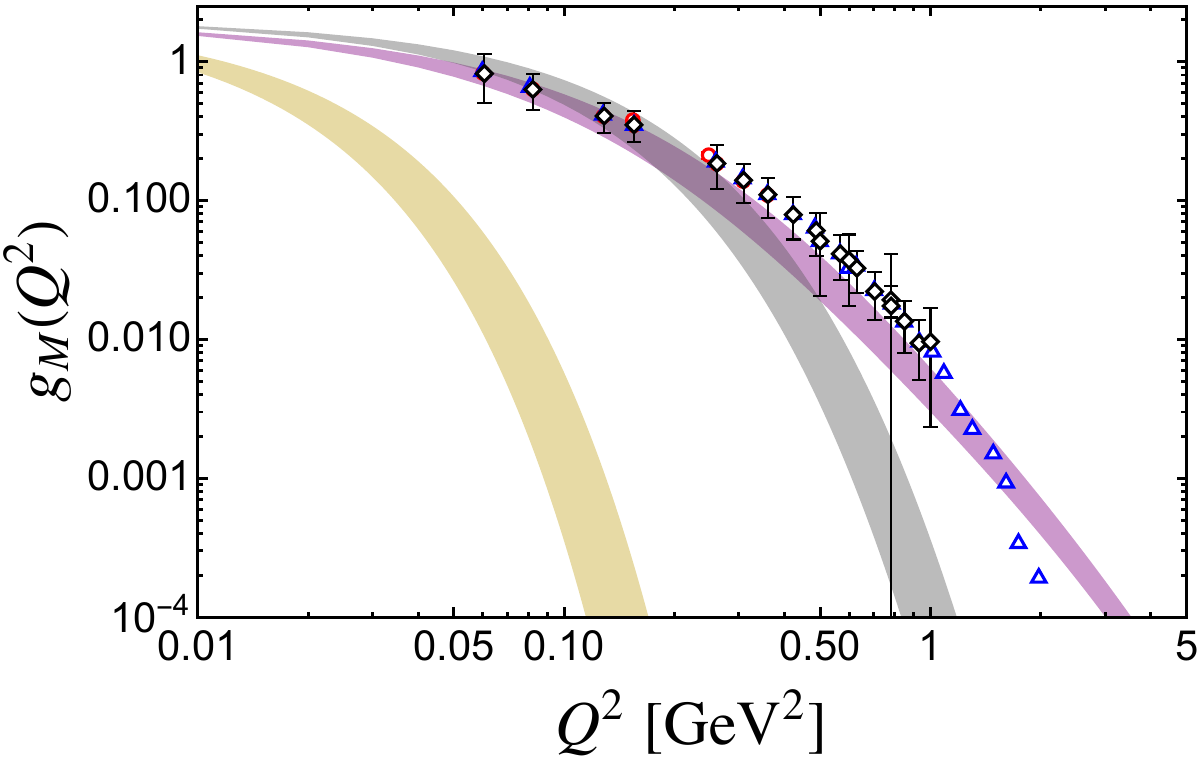}
    \includegraphics[width=0.325\textwidth]{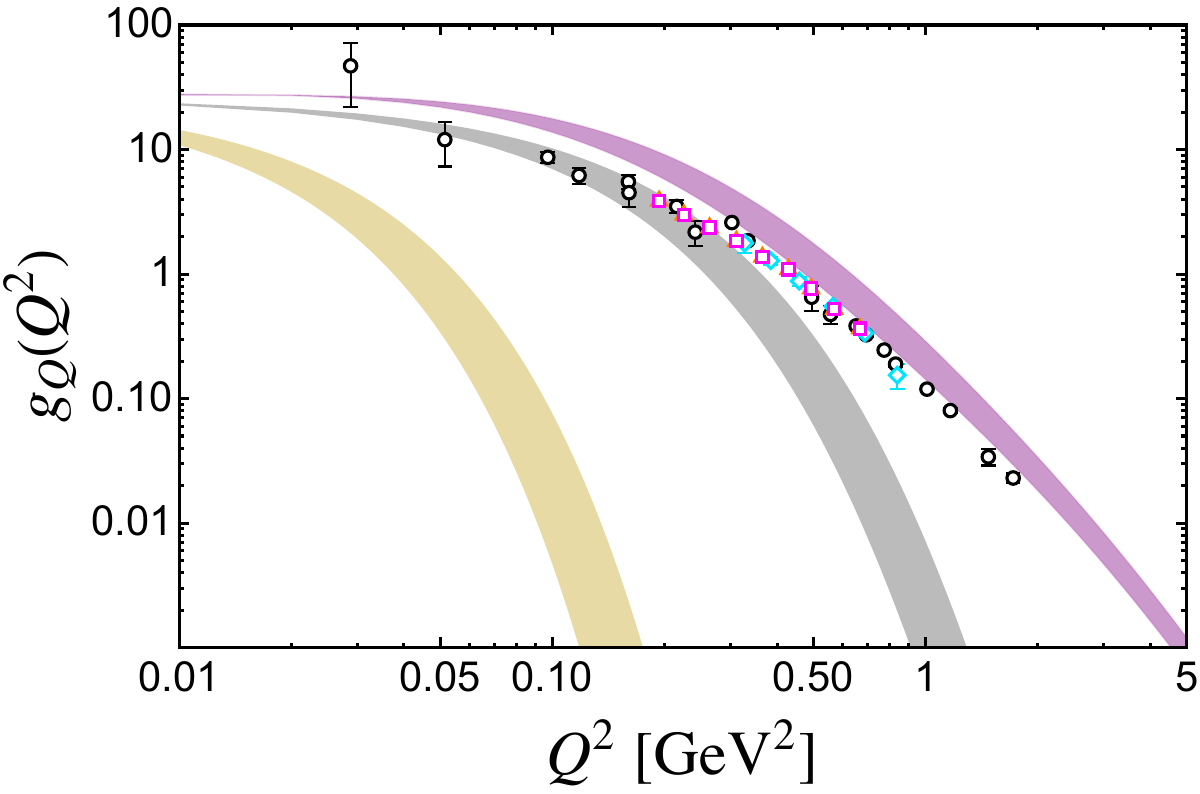}
	\caption{Comparison of the reduced electromagnetic form factors---charge, magnetic, and quadrupole---with available experimental measurements~\cite{JLABt20:2000qyq, Zhang:2011zu, The:1991eg, Benaksas:1966zz, Nikolenko:2003zq}.}
	\label{fig:EFFs}
\end{figure*}

The EMFFs of the deuteron at the level of its constituents can be expressed in terms of LF helicity transition amplitudes $I^+_{\Lambda'\Lambda} = \langle P^\prime\Lambda^\prime|J_{\rm EM}^+|P\Lambda\rangle$, where $J_{\rm EM}^+$ is the plus component of the EM current. Here $|P\Lambda\rangle$ stands for the deuteron state with momentum $P$ and helicity $\Lambda$. Note that in LF dynamics, these expressions are frame-independent.
The EMFFs are then given by~\cite{Choi:2004ww}:
\begin{equation}
\begin{aligned}
G_C =& \frac{1}{2P^+}\left[\frac{3-2\eta}{3}I^+_{++}+\frac{4\eta}{3}\frac{I^+_{+0}}{\sqrt{2\eta}}+\frac{I^+_{+-}}{3} \right], \\
G_M =& \frac{2}{2P^+}\left[I^+_{++}-\frac{I^+_{+0}}{\sqrt{2\eta}}\right], \\
G_Q =& \frac{1}{2P^+}\left[-I^+_{++}+2\frac{I^+_{+0}}{\sqrt{2\eta}}-\frac{I^+_{+-}}{\eta}\right],
\label{Eq:EMFFs}
\end{aligned}
\end{equation}
with the kinematic factor being $\eta=\frac{Q^2}{4 M_D^2}$, $I^+_{\Lambda^\prime \Lambda} = I^0_{\Lambda^\prime \Lambda}+I^3_{\Lambda^\prime \Lambda}$, and $P^+ = P^0 + P^3$.

To obtain more detailed and experimentally measurable predictions, we compute the ``reduced" nuclear form factor that removes the effects of nucleon internal structure. This is achieved by dividing the computed deuteron form factors using Eq.~\eqref{Eq:EMFFs} by the single-nucleon form factor $F_{N}(Q^2)$~\cite{Brodsky:1976mn, Brodsky:1983vf, Kobushkin:1994ed, Brodsky:1985gs}:
\begin{equation}
    g_{C,M,Q}(Q^2)\equiv \frac{G_{C,M,Q}(Q^2)}{F^2_N(Q^2/4)},
\end{equation}
where the nucleon form factor is parametrized in dipole form as $F_N(Q^2)= \left( 1+\frac{Q^2}{0.71 ~{\rm GeV}^2}\right)^{-2}$. This reduction isolates the genuine nuclear structure effects from those arising from the finite size of individual nucleons. In our framework, since the clusters have masses comparable to the nucleon mass and consist of three valence quarks at observed energy scale, we employ the nucleon form factor $F_N(Q^2)$ as a physically motivated approximation.

Figure~\ref{fig:EFFs} shows our results for the charge, magnetic, and quadrupole reduced EMFFs ($g_C$, $g_M$, and $g_Q$) compared to the available experimental data~\cite{JLABt20:2000qyq, Zhang:2011zu, The:1991eg, Benaksas:1966zz, Nikolenko:2003zq}. We observe that our approach globally describes the data. Figure~\ref{fig:EFFs} also presents the B.E. effects on EMFFs, spanning a broad range from a few MeV to $\sim 200$~MeV.

Our approach yields a magnetic moment of \( \mu_D = 0.84 \), which is in good agreement with the experimental value of 0.85~\cite{Garcon:2001sz}. However, the calculated quadrupole moment \( \mathcal{Q}_D = 0.079~\text{GeV}^{-2} \) is significantly smaller than the measured value of \( 7.34~\text{GeV}^{-2} \)~\cite{Garcon:2001sz}. This discrepancy likely arises from the absence of the \( D \)-wave component~\cite{Ericson:1985hf} in our current framework, which is known to play a crucial role in deforming the deuteron into its characteristic ``peanut" shape. To examine the \( Q^2 \)-dependence of the quadrupole form factor, we normalize \( g_Q(0) \) to the experimental value~\cite{Gutsche:2016lrz}, \( g_Q(0) = 25.83 \) in Fig.~\ref{fig:EFFs}.

We find the charge radius, 
\( \sqrt{\langle r^2_{C} \rangle} = 2.17 \pm 0.20 \) fm, consistent with the experimental value \( 2.130 \pm 0.003 \pm 0.009 \) fm~\cite{Sick:1998cvq}.
Our result for the magnetic radius, 
\( \sqrt{\langle r^2_{M} \rangle} = 1.98 \pm 0.19 \) fm, is also in good agreement with the experimental measurement of \( 1.90 \pm 0.14 \) fm~\cite{Afanasev:1998hu}.

The longitudinal momentum distribution functions (LMDFs), as one-dimensional distribution functions, describe the probability of finding a constituent carrying the fraction ($z$) of the total longitudinal momentum of the deuteron. 
Assuming the clusters are point-like particles, the LMDFs are defined via the correlator~\cite{Bacchetta:2000jk, Bacchetta:2001rb, MuldersTMDs},
\begin{equation}
	\Phi^{\Lambda}_{ij}(z) = \int \frac{{\rm d}y^-}{2 \pi} e^{\iota k^+\cdot y^-}  \langle P\Lambda \vert \bar{\phi}_j(0) U_{[0,y^-]} \phi_i(y^-) \vert P\Lambda \rangle.
	\end{equation}
The gauge link $U_{[0,y^-]}$ is unity in the LF gauge, $A^+=0$. The correlator is associated with the partonic distributions as $\Phi^{[\Gamma]}(z)=\frac{1}{2} {\rm Tr}\left(\Gamma\Phi^{\Lambda}(z) \right)$.
The use of different Dirac matrices ($\Gamma$) allows for the parametrization of the LMDFs for spin-1 bound state systems as~\cite{Bacchetta:2000jk,Bacchetta:2001rb},
\begin{equation}
\begin{aligned}
	\Phi^{[\gamma^+]}(z) &= f_1(z) + S_{LL}\,f_{1LL}(z),\\
		\Phi^{[\gamma^+\gamma_5]}(z) &= S_L\, g_{1L} (z),	\end{aligned}
        \end{equation}
with $S_L$ being the longitudinal polarization of the target, and $S_{LL}=(3\Lambda^2-2) \left( 1/6-S_L^2/2 \right)$~\cite{Kaur:2020emh}.
Note that for $\Lambda = 0$, $ S_L = 1$ and for $\Lambda = \pm 1$, $S_L = 0$.

The unpolarized ($f_1$), helicity-dependent ($g_{1L}$) and tensor-polarized ($f_{1LL}$) LMDFs of deuteron are expressed in terms of the number density of its constituents $\big(P^\Lambda_{\uparrow}(z) = \int {\rm d}^2{\bf k}_\perp \sum_{\bar{h}}\vert \Psi^\Lambda_{\uparrow \bar{h}} (z,{\bf k}^2_\perp) \vert^2$, $P^\Lambda_{\downarrow}(z) = \int {\rm d}^2{\bf k}_\perp \sum_{\bar{h}}\vert \Psi^\Lambda_{\downarrow \bar{h}}(z,{\bf k}^2_\perp)\vert^2 \big)$ with helicity up $(\uparrow)$ or down $(\downarrow)$ in a deuteron with helicity polarization $\Lambda$~\cite{Kaur:2020emh},
	\begin{equation}
    \begin{aligned}
		f_1(z) &= \frac{2}{3}\left(P^1_\uparrow(z) + P^{-1}_\uparrow(z) + P^{0}_\uparrow(z)\right),  \\
		g_{1L}(z) &= P^1_\uparrow(z) - P^{1}_\downarrow(z), \label{eq:g1LN}\\
		f_{1LL}(z) &= 2 P^0_\uparrow(z) - \left(P^{1}_\uparrow(z) + P^{-1}_\uparrow(z)\right).
        \end{aligned}
	\end{equation}

	\begin{figure}[hbt!]
	\centering		\includegraphics[width=0.36\textwidth]{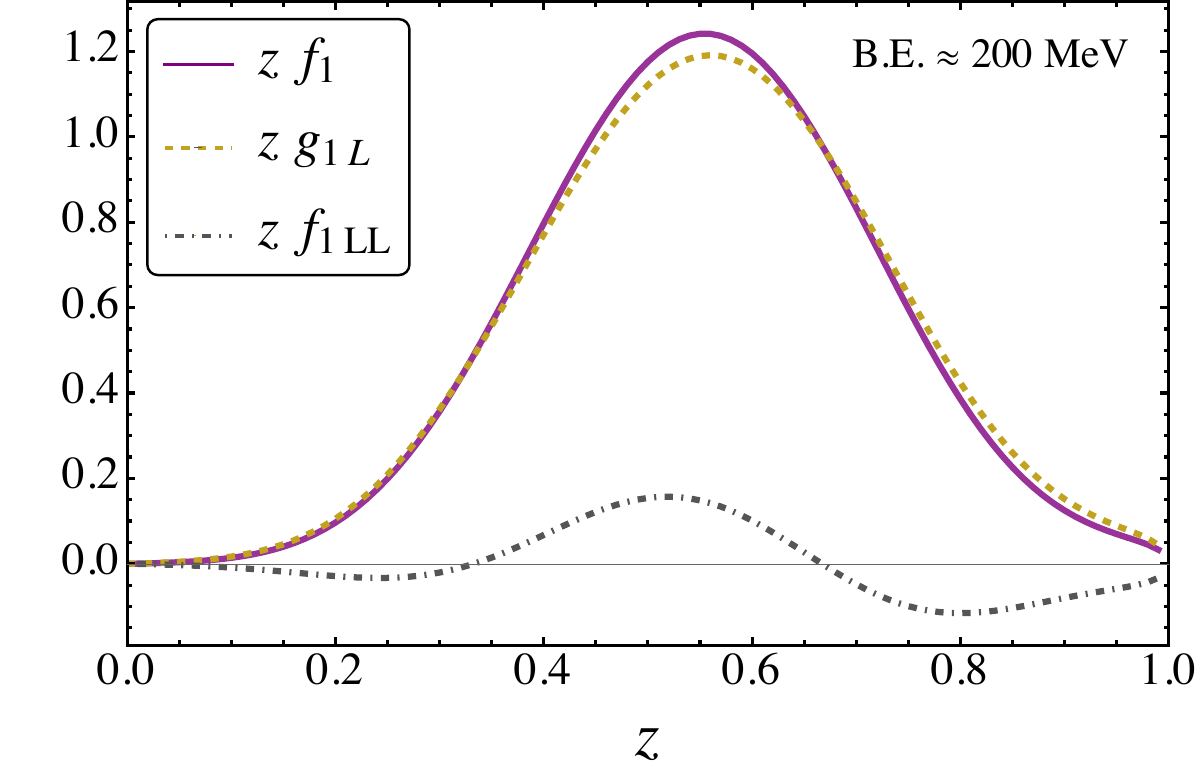}
	\caption{The longitudinal momentum distribution functions of a cluster inside the deuteron.}
	\label{fig:Nucleon-PDFs}
\end{figure}
    
In Fig.~\ref{fig:Nucleon-PDFs}, we present the LMDFs of a cluster inside the deuteron.
The distribution function $f_1$
gives the probability of finding an unpolarized cluster-constituent carrying a fraction $z$ of the longitudinal momentum of an unpolarized deuteron. On the other hand, $g_{1L}$ represents the probability of a longitudinally polarized cluster carrying the fraction $z$ of the longitudinal momentum of a longitudinally-polarized deuteron. The distribution function $f_{1LL}$ is unique to spin-1 QCD bound state systems, representing the distribution of unpolarized constituents inside a tensor-polarized deuteron. 
It is noteworthy that our distribution functions satisfy the sum rules~\cite{Kaur:2020emh}:
$\int_0^1 {\rm d} z  f_1(z) = 1 \,\text{and}\, \int_0^1 {\rm d}z f_{1LL}(z)=0.$

	\begin{figure*}[hbt!]
	\centering
\includegraphics[width=0.31\textwidth]{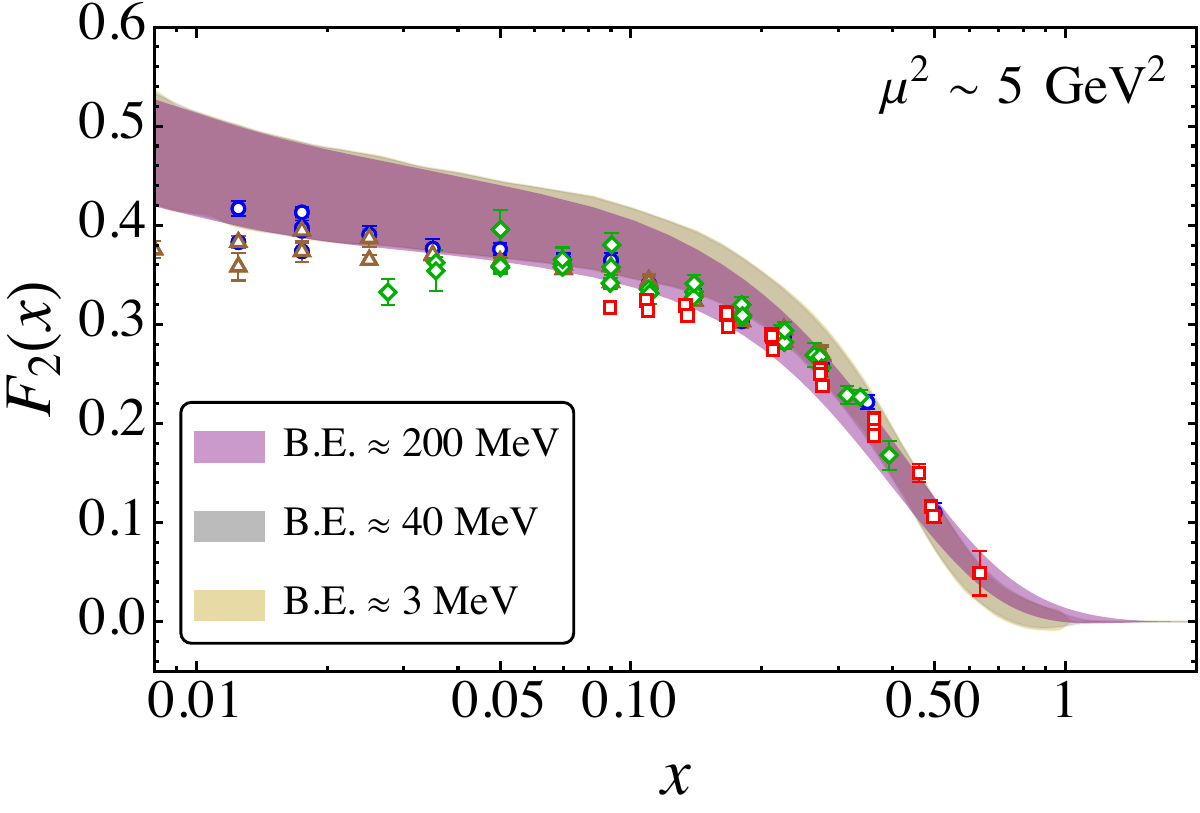}
    \includegraphics[width=0.32\textwidth]{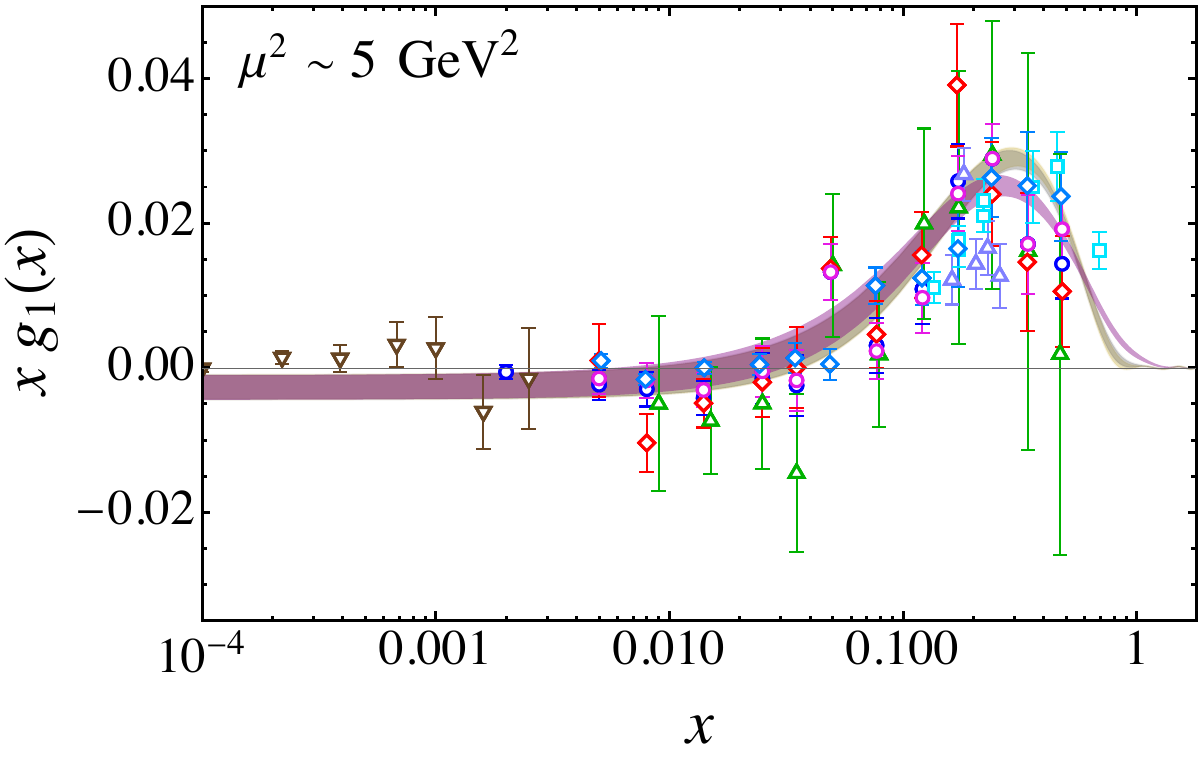}
    \includegraphics[width=0.32\textwidth]{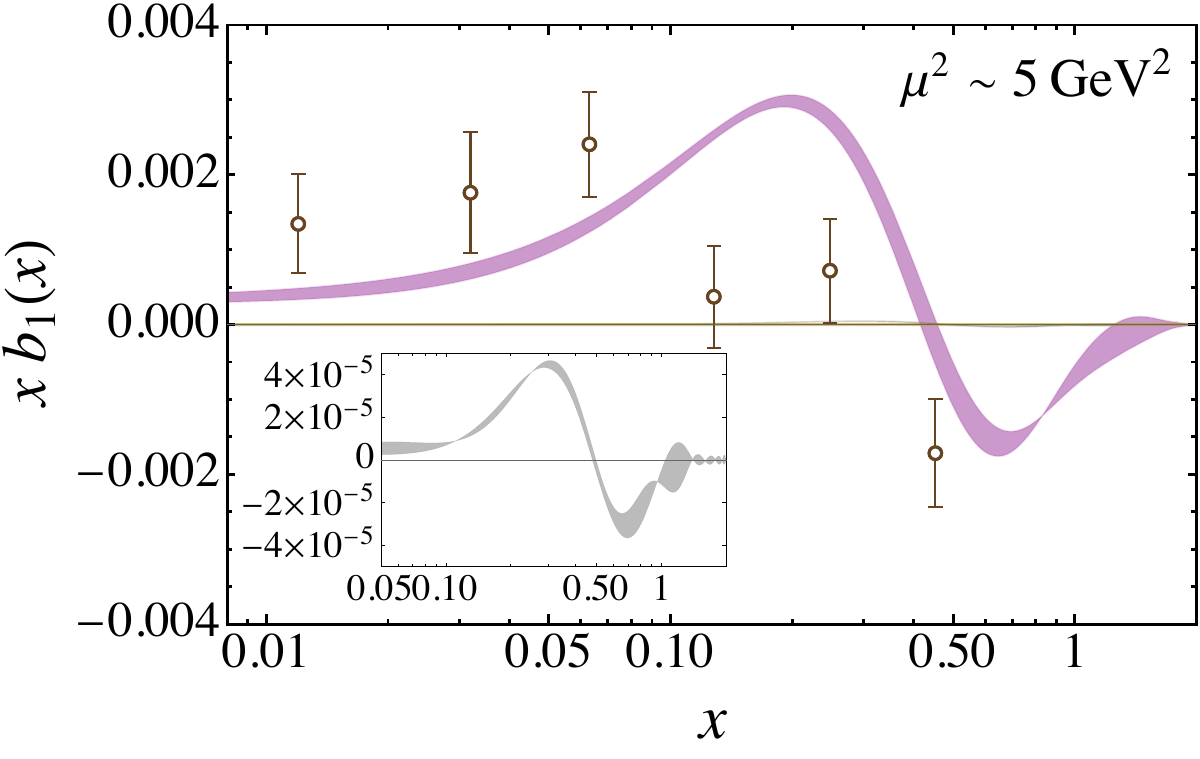}
	\caption{Comparison of the unpolarized, helicity, and tensor-polarized structure functions of deuteron at scale $\mu^2\sim5$ GeV$^2$ with experimental data~\cite{NewMuon:1995zeb, NewMuon:1996fwh, Gabbert:2008zz, SpinMuon:1998eqa, SpinMuon:1993gcv, SpinMuon:1995svc, SpinMuon:1998eqa, SpinMuon:1999udj, HERMES:2006jyl, E143:1998hbs, COMPASS:2005xxc, HERMES:2005pon}. Note that for B.E. $\approx$ 3~MeV,  $x b_1(x)$ vanishes and is thus not visible.}
	\label{fig:Structure-functions}
\end{figure*}

    The structure functions of the deuteron encode its quark-gluon substructure across different polarization states.
    The fundamental relationship between the structure function $\mathrm{SF}^D(x,Q^2)$ and the quark distributions $\mathcal{F}_q^D(x,Q^2)$ is given by
    \begin{equation}
\begin{aligned}
\mathrm{SF}^D(x,Q^2) = x \sum_q e_q^2 \mathcal{F}_q^D(x,Q^2),
\label{eq:SF}
\end{aligned}
\end{equation}
where $x$ denotes the momentum fraction carried by the struck quark within a cluster, and $e_q$ represents the quark charge (e.g., $e_u = +2/3$, $e_d = -1/3$).

The deuteron's PDF $\mathcal{F}_q^D$ factorizes into the convolution of two key components: (i) the LMDF of two clusters within the deuteron, and (ii) the quark PDF $\mathcal{F}_q^N$ within each cluster. Mathematically, this reads
\begin{align}
\mathcal{F}_q^D(x,Q^2) = \frac{1}{2} \sum_{N} \int_x^1 \frac{dz}{z} \, \mathcal{\mathrm{LMDF}}(z) \, \mathcal{F}_q^N\left(\frac{x}{z}, Q^2\right),
\label{eq:PDF}
\end{align}
where $\mathcal{F}_q^N$ is the ``cluster" PDF adapted from the NNPDF global fits at $\mu^2\sim 5$~GeV$^2$~\cite{NNPDF:2017mvq}.

The unpolarized $F_2$ involves unpolarized LMDFs paired with unpolarized quark densities, while the helicity-sensitive $g_1$ employs polarized cluster distributions and quark helicity densities. The tensor-polarized $b_1$ emerges from tensor-polarized LMDFs coupled to unpolarized quark PDFs. 

Figure~\ref{fig:Structure-functions} shows good agreement between our deuteron structure functions and the comprehensive experimental data~\cite{NewMuon:1995zeb, NewMuon:1996fwh, Gabbert:2008zz, SpinMuon:1998eqa, SpinMuon:1993gcv, SpinMuon:1995svc, SpinMuon:1998eqa, SpinMuon:1999udj, HERMES:2006jyl, E143:1998hbs, COMPASS:2005xxc, HERMES:2005pon}. The remarkable consistency across the entire $x$-range validates our cluster-based approach. We also examine the dependence of structure functions on the B.E.. The results indicate that the unpolarized ($F_2$) and helicity ($xg_1$) structure functions are largely insensitive to the variation in B.E., particularly at small-$x$.

The tensor-polarized structure function $b_1$ probes novel QCD dynamics in the deuteron. Figure~\ref{fig:Structure-functions} (right panel) shows our prediction compared to the HERMES data~\cite{HERMES:2005pon}, with good qualitative agreement. Notably, $b_1$ vanishes at small B.E., where the deuteron behaves like a loosely bound system of two nucleons. The inset in the right panel in Fig.~\ref{fig:Structure-functions} reveals the tininess of $b_1$ even at B.E. $\approx$ 40 MeV. In fact, previous studies have shown that in a pure $S$-wave configuration, $b_1$ is expected to vanish, and even with a $D$-wave, the observed data could not be explained~\cite{Cosyn:2017fbo,Kumano:2024fpr}. Our results indicate that hidden-color degrees of freedom, which become more significant at higher B.E., play a crucial role in generating a nonzero $b_1$, offering a possible explanation for the HERMES data.

Our computed value of the first moment of \( b_1(x) \) at $5$~GeV\(^2\),  
$
\int_{0.02}^{0.85} {\rm d}x \, b_1(x) = (0.36 \pm 0.03) \times 10^{-2}
$  
agrees with the HERMES data,  
\((0.35 \pm 0.10_{\rm stat} \pm 0.18_{\rm sys}) \times 10^{-2}\)~\cite{HERMES:2005pon}.

	\begin{figure}[hbt!]
	\centering
	\includegraphics[width=0.35\textwidth]{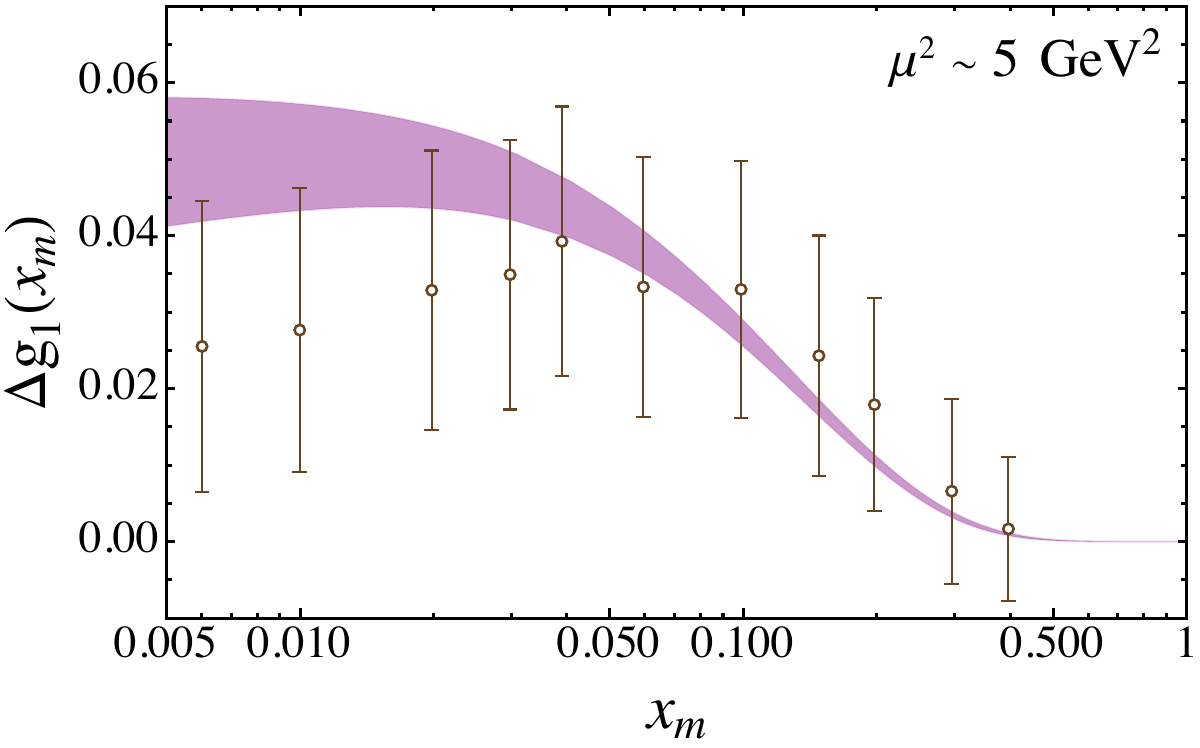}
	\caption{Quark contribution to the deuteron's spin versus $x_m$ compared with the experimental data~\cite{SpinMuon:1993gcv}.}
	\label{fig:moments-g1}
\end{figure}

Figure~\ref{fig:moments-g1} shows the $x$-moment of $g_1$, i.e., $\Delta g_1(x_m)= \int^1_{x_m} {\rm d}x g_1(x)$. Our results show good compatibility with the experimental data~\cite{SpinMuon:1993gcv}.

{\it Conclusions and Outlook}.---The deuteron, a cornerstone of nuclear physics, provides crucial insights into the fundamental forces shaping matter. This work presents the first attempt of calculating the internal structure of the deuteron incorporating hidden color degrees of freedom to reveal non-nucleonic contributions in its internal dynamics. We solved for the deuteron mass and wavefunctions using two Schr\"odinger-like equations, light-front holography and the 't~Hooft equation, for the discussion of color configurations in QCD. These solutions enabled calculations of key observables, including electromagnetic form factors and structure functions. 

The overall quality of the description is noteworthy, particularly given the simplicity of our approach and the  wide range of data sets it successfully reproduces. The agreement suggests that the essential physics of the deuteron's structure is well-captured by our color-cluster convolution approach.
In particular, our results indicate a possible explanation for the HERMES observations of the tensor structure function $b_1(x)$ with hidden color degrees of freedom.

We plan to include the $D$-wave contribution in the future work to analyze the deuteron quadrupole moment more extensively and incorporate the QCD evolution equation to quantify the relative contributions of singlet-singlet and octet-octet components in the light-front wavefunction more explicitly. We think that the insights gained from these analyses will enhance our understanding of the internal dynamics of the deuteron as well as the heavier nuclei with upcoming experiments at Jefferson Lab and future EICs.

{\it Acknowledgments}.---We thank Shunzo Kumano and Tobias Frederico for insightful and fruitful discussions. SK is supported by Research Fund for International Young
Scientists, Grant No.~12250410251, from the National Natural Science Foundation of China (NSFC), and China Postdoctoral Science
Foundation (CPSF), Grant No.~E339951SR0. CM is supported by new faculty start up funding by the Institute of Modern Physics, Chinese Academy of Sciences, Grant No.~E129952YR0.  CM also thanks the Chinese Academy of Sciences Presidents International Fellowship Initiative for support via Grants No.~2021PM0023. XZ is supported by new faculty start-up funding by the Institute of Modern Physics, 
Chinese Academy of Sciences, by the Key Research Program of Frontier Sciences, 
Chinese Academy of Sciences, Grant No.~ZDBS-LY-7020, by the Natural Science Foundation 
of Gansu Province, China, Grant No.~20JR10RA067, by the Foundation for Key Talents 
of Gansu Province, by the Central Funds Guiding the Local Science and Technology 
Development of Gansu Province, Grant No.~22ZY1QA006, by the International Partnership 
Program of the Chinese Academy of Sciences, Grant No.~016GJHZ2022103FN, by the Strategic 
Priority Research Program of the Chinese Academy of Sciences, Grant No.~XDB34000000, 
and by the National Natural Science Foundation of China under Grant No.~12375143. 
CJ is supported in part by the U.S. Department of Energy
(Grant No.~DE-FG02-03ER41260) and acknowledges the National Energy Research Scientific Computing Center (NERSC) supported by the Office of Science of the U.S. Department of Energy under Contract No.~DE-AC02-05CH11231.

\bibliography{ref.bib}
\end{document}